\documentclass{aa}
\usepackage{graphicx}
\usepackage{epsf}
\begin{document}
\thesaurus{02            % A&A section 2, cosmology
              (12.03.4;  % cosmology: theory
               12.04.1;  % cosmology: dark matter
               12.12.1;  % cosmology: large-scale structure of universe
               11.07.1;  % galaxies: general
               11.17.3)} % galaxies: quasars: general
\title{Two-point correlation functions on the light cone: \\
         testing theoretical predictions against N-body simulations}

\author{Takashi Hamana  \inst{1}, St\'ephane Colombi \inst{1,2},
 \and Yasushi Suto \inst{3,4}}

\offprints{T. Hamana}

\institute{Institut d'Astrophysique de Paris, CNRS, 
98bis Boulevard Arago, F 75014 Paris, France\\
email: hamana@iap.fr, colombi@iap.fr
\and {\cal NIC} (Numerical Investigations in Cosmology) Group, CNRS
\and
Department of Physics, University of Tokyo, Tokyo 113-0033, Japan\\
email: suto@phys.s.u-tokyo.ac.jp
\and
Research Center for the Early Universe (RESCEU),
School of Science, University of Tokyo, Tokyo 113-0033, Japan\\
}
\titlerunning{Two-point correlation functions on the light cone}
\authorrunning{Hamana, Colombi,  \& Suto}

\date{Received July 3, 2000 ; accepted 30 November, 2000}
\maketitle
\begin{abstract}
  We examine the light-cone effect on the two-point correlation
  functions using numerical simulations for the first time.
  Specifically, we generate several sets of dark matter particle
  distributions on the light-cone up to $z=0.4$ and $z=2$ over the
  field-of-view of $\pi$ degree$^2$ from cosmological N-body
  simulations. Then we apply the selection function to the dark matter
  distribution according to the galaxy and QSO luminosity functions.
  Finally we compute the two-point correlation functions on the
  light-cone both in real and in redshift spaces using the pair-count
  estimator and compare with the theoretical predictions.  We find
  that the previous theoretical modeling for nonlinear gravitational
  evolution, linear and nonlinear redshift-distortion, and the
  light-cone effect including the selection function is in good
  agreement with our numerical results, and thus is an accurate and
  reliable description of the clustering in the universe on the
  light-cone.  
\end{abstract}

%%%%%%%%%%%%%%%%%%%%%%%%%%%%%%%%%%%%%%%%%%%%%%%%%%%%%%%%%%%%%%%%%%%%%%%%%

\section{Introduction}

In the proper understanding of on-going redshift surveys of galaxies and
quasars, in particular the Two-degree Field (2dF) and the Sloan Digital
Sky Survey (SDSS), it is essential to establish a theory of cosmological
statistics on the light cone. This project has been undertaken in a
series of our previous papers (Matsubara, Suto, \& Szapudi 1997;
Yamamoto \& Suto 1999; Nishioka \& Yamamoto 1999; Suto et~al. 1999;
Yamamoto, Nishioka, \& Suto 2000; Suto, Magira \& Yamamoto 2000). Those
papers have formulated the light-cone statistics in a rigorous manner,
described approximations to model the clustering evolution in the
redshift space, and presented various predictions in canonical cold dark
matter (CDM) universes.  Their predictions, however, have not yet been
tested quantitatively, for instance, against numerical simulations. This
is not surprising since it is fairly a demanding task to construct a
reliable sample extending over the light-cone from the conventional
simulation outputs at a specified redshift, $z$.

In the present paper, we examine, for the first time, the validity and
limitation of the above theoretical framework to describe the
cosmological light-cone effect against the mock catalogues on the
light-cone. Such catalogues from cosmological $N$-body simulations
have been originally constructed for the study of the weak lensing
statistics (Hamana et al. 2000, in preparation). Applying the same
technique (\S\ref{subsec:lcfromnbody}), we generate a number of
different realizations for the light-cone samples up to $z=0.4$ and
$z=2$, evaluate the two-point correlation functions directly, and
compare with the theoretical predictions.

\section{Predictions of two-point correlation functions on the light cone}

In order to predict quantitatively the two-point statistics of objects
on the light cone, one must take account of (i) nonlinear
gravitational evolution, (ii) linear redshift-space distortion, (iii)
nonlinear redshift-space distortion, (iv) weighted averaging over the
light-cone, (v) cosmological redshift-space distortion due to the
geometry of the universe, and (vi) object-dependent clustering bias.
The effect (v) comes from our ignorance of the correct cosmological
parameters, and (vi) is rather sensitive to the objects which one has
in mind. Thus the latter two effects will be discussed in a separate
paper, and we focus on the effects of (i) $\sim$ (iv) throughout the
present paper.

Nonlinear gravitational evolution of mass density fluctuations is now
well understood, at least for two-point statistics. In practice, we
adopt an accurate fitting formula (Peacock \& Dodds 1996) for the
nonlinear power spectrum $P^{\rm R}_{\rm nl}(k,z)$ in terms of its
linear counterpart.

Then the nonlinear power spectrum in redshift space is given as
%%%%%%%%%%%%%%%%%%%%%%%%%%%%%%%%%%%%%%%%%%%%%%%%%%%%%%%%%%%%%%%%%%%%%%%%%%
\begin{equation}
\label{eq:pknls}
  P^{\rm S}_{\rm nl}(k,\mu)
=P^{\rm R}_{\rm nl}(k,z)[1+\beta\mu^2]^2 D_{\rm vel}[k\mu\sigma_P] ,
\end{equation}
%%%%%%%%%%%%%%%%%%%%%%%%%%%%%%%%%%%%%%%%%%%%%%%%%%%%%%%%%%%%%%%%%%%%%%%%%%
where $k$ is the comoving wavenumber, and $\mu$ is the direction
cosine in $k$-space.  The second factor in the right-hand-side comes
from the linear redshift-space distortion (Kaiser 1987), and the last
factor is a phenomenological correction for non-linear velocity
effect.  In the above, we introduce
%%%%%%%%%%%%%%%%%%%%%%%%%%%%%%%%%%%%%%%%%%%%%%%%%%%%%%%%%%%%%%%%%%%%%%%%%%%%
\begin{eqnarray} 
\label{eq:betaz}
   \beta(z) \equiv {1\over b(z)} \frac{d\ln D(z)}{d\ln a} ,
\end{eqnarray}
%%%%%%%%%%%%%%%%%%%%%%%%%%%%%%%%%%%%%%%%%%%%%%%%%%%%%%%%%%%%%%%%%%%%%%%%%%%%
where $D(z)$ is the gravitational growth rate of the linear density
fluctuations, $a$ is the cosmic scale factor, and the density
parameter, the cosmological constant, and the Hubble parameter at
redshift $z$ are related to their present values respectively as
%%%%%%%%%%%%%%%%%%%%%%%%%%%%%%%%%%%%%%%%%%%%%%%%%%%%%%%%%%%%%%%%%%%%%%%%%%%%
\begin{eqnarray} 
   \Omega(z) &=& \left[{H_0 \over H(z)}\right]^2 \, (1+z)^3 \Omega_0 ,\\
 \lambda(z) &=& \left[{H_0 \over H(z)}\right]^2 \, \lambda_0 , \\
   H(z) &=& H_0\sqrt{\Omega_0 (1 + z)^3 + 
                  (1-\Omega_0-\lambda_0) (1 + z)^2 + \lambda_0} .
\end{eqnarray}
%%%%%%%%%%%%%%%%%%%%%%%%%%%%%%%%%%%%%%%%%%%%%%%%%%%%%%%%%%%%%%%%%%%%%%%%%%%%
We assume that the pair-wise velocity distribution in real space is
approximated by
%%%%%%%%%%%%%%%%%%%%%%%%%%%%%%%%%%%%%%%%%%%%%%%%%%%%%%%%%%%%%%%%%
\begin{equation}
\label{eq:expveldist}
  f_v(v_{12}) = {1 \over \sqrt{2}{\hbox {$\sigma_{\scriptscriptstyle
         {\rm P}}$}}} \exp\left(-{\sqrt{2}|v_{12}| \over {\hbox
         {$\sigma_{\scriptscriptstyle {\rm P}}$}}} \right) ,
\end{equation}
%%%%%%%%%%%%%%%%%%%%%%%%%%%%%%%%%%%%%%%%%%%%%%%%%%%%%%%%%%%%%%%%%
with ${\hbox {$\sigma_{\scriptscriptstyle {\rm P}}$}}$ being the
1-dimensional pair-wise peculiar velocity dispersion.  In this case
the damping term in Fourier space, $D_{\rm vel}[k\mu\sigma_P]$, is given by
%%%%%%%%%%%%%%%%%%%%%%%%%%%%%%%%%%%%%%%%%%%%%%%%%%%%%%%%%%%%%%%%%%%%%%%%%%
\begin{equation}
\label{eq:damping}
  D_{\rm vel}[k\mu\sigma_{\scriptscriptstyle {\rm P}}]
=\frac{1}{1+\kappa^2\mu^2} ,
\end{equation}
%%%%%%%%%%%%%%%%%%%%%%%%%%%%%%%%%%%%%%%%%%%%%%%%%%%%%%%%%%%%%%%%%%%%%%%%%%
where
%%%%%%%%%%%%%%%%%%%%%%%%%%%%%%%%%%%%%%%%%%%%%%%%%%%%%%%%%%%%%%%%%%%%%%%%%%
\begin{equation}
\kappa(z)= \frac{k(1+z)\sigma_{\scriptscriptstyle {\rm P}}(z)}
{\sqrt{2}H(z)}.
\end{equation}
%%%%%%%%%%%%%%%%%%%%%%%%%%%%%%%%%%%%%%%%%%%%%%%%%%%%%%%%%%%%%%%%%%%%%%%%%%
Note that this expression is equivalent to that in Magira et al.
(2000) but written in terms of the physical velocity units.

On large scales, $\sigma_{\scriptscriptstyle {\rm P}}(z)$ can be well
approximated by a fitting formula proposed by Mo, Jing \& B\"{o}rner
(1997):
%%%%%%%%%%%%%%%%%%%%%%%%%%%%%%%%%%%%%%%%%%%%%%%%%%%%%%%%%%%%%%%%%%%
\begin{eqnarray}
\label{eq:sigma_p}
  \sigma_{\rm\scriptscriptstyle P,MJB}^2(z) &\equiv& 
    \frac{\Omega(z)H^2(z)}{(1+z)^2}
   \left[1-\frac{1+z}{D^2(z)}
   \int_z^\infty \frac{D^2(z')}{(1+z')^2}dz'\right] \cr
&\qquad& \hspace*{2cm} \times
   \int_0^{\infty}\frac{dk}{k}
 \frac{\Delta^2_{\rm\scriptscriptstyle NL}(k,z)}{k^2} .
\end{eqnarray}
%%%%%%%%%%%%%%%%%%%%%%%%%%%%%%%%%%%%%%%%%%%%%%%%%%%%%%%%%%%%%%%%%%%

%%%%%%%% FIG 1 %%%%%%%%%%%%%%%%%%%%%%%%%%%%%%%%%%%%%%%%%%%%%%%%%%%%%%%
\begin{figure*}
\begin{center}
   \leavevmode \epsfxsize=12cm \epsfbox{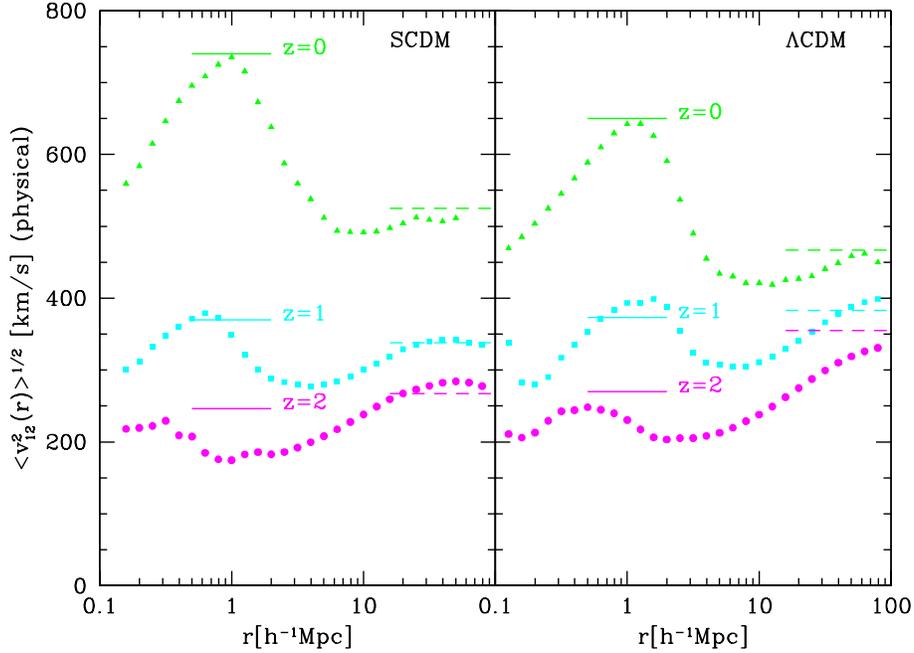}
\end{center}
\caption{Pairwise peculiar velocity dispersions of dark matter
  particles at $z=0$, 1 and 2. Dashed lines indicate the values
  predicted from the formula of MJB (eq.[\ref{eq:sigma_p}]), while
  solid lines indicate our adopted fit (eq.[\ref{eq:sigma_pfit}]). }
\label{fig:pairvrms}
\end{figure*}
%%%%%%%%%%%%%%%%%%%%%%%%%%%%%%%%%%%%%%%%%%%%%%%%%%%%%%%%%%%%%%%%%%%%%

We compute the pairwise velocity dispersion of particles in $N$-body
simulations (see \S 3.1) both for SCDM and $\Lambda$CDM,
whose parameters are summarized in Table \ref{tab:parameters}, to test
the accuracy of the fitting formula of the pairwise velocity
dispersion, eq.~(\ref{eq:sigma_p}).  The measured velocity dispersions
at $z=0$, 1 and 2 are shown in Figure \ref{fig:pairvrms}.  The dotted
lines in Figure \ref{fig:pairvrms} indicate predictions of
eq.~(\ref{eq:sigma_p}) integrated over the wavenumbers existing in our
$N$-body simulations.  The analytical model predictions agree with our
data within a $10\%$ accuracy at the large separations.  This level of
agreement is as good as that found originally by Mo et al.~(1997).
Nevertheless since we are mainly interested in the scales around
$1h^{-1}$Mpc, we adopt the following fitting formula throughout the
analysis below which better approximates the small-scale dispersions
in physical units:
%%%%%%%%%%%%%%%%%%%%%%%%%%%%%%%%%%%%%%%%%%%%%%%%%%%%%%%%%%%%%%%%%%%
\begin{eqnarray}
\label{eq:sigma_pfit}
  \sigma_{\rm\scriptscriptstyle P}(z) \sim \left\{ 
      \begin{array}{ll}
         740 (1+z)^{-1} {\rm km/s} & \mbox{for SCDM model} \\
         650 (1+z)^{-0.8} {\rm km/s} & \mbox{for $\Lambda$CDM model}.
      \end{array}
   \right.
\end{eqnarray}
%%%%%%%%%%%%%%%%%%%%%%%%%%%%%%%%%%%%%%%%%%%%%%%%%%%%%%%%%%%%%%%%%%%

Integrating equation (\ref{eq:pknls}) over $\mu$, one obtains the
direction-averaged power spectrum in redshift space:
%%%%%%%%%%%%%%%%%%%%%%%%%%%%%%%%%%%%%%%%%%%%%%%%%%%%%%%%%%%%%%%%%%%%%%%%%%
\begin{equation}
 \frac{ P^{\rm S}_{\rm nl}(k,z)}{P^{\rm R}_{\rm nl}(k,z)} =
A(\kappa)+{2\over3}\beta(z) B(\kappa)+{1\over 5}\beta^2(z) 
C(\kappa)
\label{eq:PSnl}
\end{equation}
%%%%%%%%%%%%%%%%%%%%%%%%%%%%%%%%%%%%%%%%%%%%%%%%%%%%%%%%%%%%%%%%%%%%%%%%%%
where
%%%%%%%%%%%%%%%%%%%%%%%%%%%%%%%%%%%%%%%%%%%%%%%%%%%%%%%%%%%%%%%%%%
\begin{eqnarray}
 A(\kappa) &=& {{\arctan}(\kappa) \over \kappa}, \\
  B(\kappa) &=& {3\over\kappa^2}
    \left[1-{{\arctan}(\kappa) \over \kappa}\right] ,\\
  C(\kappa) &=& {5\over3\kappa^2}\left[1-{3\over\kappa^2} +
{3 \, {\arctan}(\kappa) \over \kappa^3} \right] .
\end{eqnarray}
%%%%%%%%%%%%%%%%%%%%%%%%%%%%%%%%%%%%%%%%%%%%%%%%%%%%%%%%%%%%%%%%%%

Adopting those approximations, the direction-averaged correlation
functions on the light-cone are finally computed as
%%%%%%%%%%%%%%%%%%%%%%%%%%%%%%%%%%%%%%%%%%%%%%%%%%%%%%%%%%%%%%%%%%%
\begin{eqnarray}
\label{eq:lccrdximom}
    \xi^{\rm\scriptscriptstyle {LC}}(x_{\rm s}) 
&=& {
   {\displaystyle 
     \int_{z_{\rm min}}^{z_{\rm max}} dz 
     {dV_{\rm c} \over dz} ~[\phi(z)n_0(z)]^2
    \xi(x_{\rm s};z)
    }
\over
    {\displaystyle
     \int_{z_{\rm min}}^{z_{\rm max}} dz 
     {dV_{\rm c} \over dz}  ~[\phi(z)n_0(z)]^2
     }
} ,
\end{eqnarray}
%%%%%%%%%%%%%%%%%%%%%%%%%%%%%%%%%%%%%%%%%%%%%%%%%%%%%%%%%%%%%%%%%%%
where $z_{\rm min}$ and $z_{\rm max}$ denote the redshift range of the
survey, and
%%%%%%%%%%%%%%%%%%%%%%%%%%%%%%%%%%%%%%%%%%%%%%%%%%%%%%%%%%%%%%%%%%
\begin{eqnarray}
\xi(x_{\rm s};z) \equiv 
{1 \over 2\pi^2}\int_0^\infty
 P^{\rm S}_{\rm nl}(k,z) {\sin kx_{\rm s} \over kx_{\rm s}} k^2\,dk .
\end{eqnarray}
%%%%%%%%%%%%%%%%%%%%%%%%%%%%%%%%%%%%%%%%%%%%%%%%%%%%%%%%%%%%%%%%%%

Throughout the present analysis, we assume a standard Robertson --
Walker metric of the form:
%%%%%%%%%%%%%%%%%%%%%%%%%%%%%%%%%%%%%%%%%%%%%%%%%%%%%%%%%%%%%%%%%%%
\begin{equation}
ds^2 = -dt^2 + a(t)^2 
\{ d\chi^2 + S_K(\chi)^2 [d\theta^2 + \sin^2\theta d\phi^2 ] \} ,
\end{equation}
%%%%%%%%%%%%%%%%%%%%%%%%%%%%%%%%%%%%%%%%%%%%%%%%%%%%%%%%%%%%%%%%%%%
where $S_K(\chi)$  is determined by the sign of the curvature $K$ as
%%%%%%%%%%%%%%%%%%%%%%%%%%%%%%%%%%%%%%%%%%%%%%%%%%%%%%%%%%%%%%%%%%%
\begin{equation}
 S_K(\chi) = 
  \left\{ 
      \begin{array}{ll}
         \sin{(\sqrt{K}\chi)}/\sqrt{K} & \mbox{$(K>0)$} \\
         \chi & \mbox{$(K=0)$} \\
         \sinh{(\sqrt{-K}\chi)}/\sqrt{-K} & \mbox{$(K<0)$} 
      \end{array}
   \right. 
 \\ \nonumber
\end{equation}
%%%%%%%%%%%%%%%%%%%%%%%%%%%%%%%%%%%%%%%%%%%%%%%%%%%%%%%%%%%%%%%%%%%
The radial comoving distance $\chi(z)$ is computed by
%%%%%%%%%%%%%%%%%%%%%%%%%%%%%%%%%%%%%%%%%%%%%%%%%%%%%%%%%%%%%%%%%%%
\begin{eqnarray} 
\chi(z) &=& \int_t^{t_0} {dt \over a(t)}
= { 1 \over a_0 } \int_0^z {d z \over H(z)} .
\end{eqnarray}
%%%%%%%%%%%%%%%%%%%%%%%%%%%%%%%%%%%%%%%%%%%%%%%%%%%%%%%%%%%%%%%%%%%
In our definition, $K$ is not normalized to $\pm1$ and $0$, but rather
written in terms of the scale factor at present, $a_0$, the Hubble
constant, $H_0$, the density parameter, $\Omega_0$ and the dimensionless
cosmological constant, $\lambda_0$:
%%%%%%%%%%%%%%%%%%%%%%%%%%%%%%%%%%%%%%%%%%%%%%%%%%%%%%%%%%%%%%%%%%%
\begin{equation}
K = a_0^2 H_0^2 (\Omega_0 + \lambda_0 -1) .
\end{equation}
%%%%%%%%%%%%%%%%%%%%%%%%%%%%%%%%%%%%%%%%%%%%%%%%%%%%%%%%%%%%%%%%%%%

The comoving angular diameter distance $D_c(z)$ at redshift $z$ is
equivalent to $S^{-1}(\chi(z))$, and, in the case of $\lambda_0=0$, is
explicitly given by Mattig's formula:
%%%%%%%%%%%%%%%%%%%%%%%%%%%%%%%%%%%%%%%%%%%%%%%%%%%%%%%%%%%%%%%%%%%
\begin{equation}
D_c(z) = {1 \over a_0 H_0} {z \over 1+z}
{1+z+\sqrt{1+\Omega_0z} \over 1 +\Omega_0 z/2 +\sqrt{1+\Omega_0z}} .
\end{equation}
%%%%%%%%%%%%%%%%%%%%%%%%%%%%%%%%%%%%%%%%%%%%%%%%%%%%%%%%%%%%%%%%%%%
Then $dV_{\rm c}/dz$, the comoving volume element per unit solid
angle, is explicitly given as
%%%%%%%%%%%%%%%%%%%%%%%%%%%%%%%%%%%%%%%%%%%%%%%%%%%%%%%%%%%%%%%%%%%
\begin{eqnarray}
{dV_{\rm c} \over dz} &=& S_K^2(\chi) {d\chi \over dz} \cr
&=& { S_K^2(\chi)\over
H_0 \sqrt{\Omega_0 (1 + z)^3 + (1-\Omega_0-\lambda_0) (1 + z)^2 +
  \lambda_0} } .
\end{eqnarray}
%%%%%%%%%%%%%%%%%%%%%%%%%%%%%%%%%%%%%%%%%%%%%%%%%%%%%%%%%%%%%%%%%%%

\section{Evaluating two-point correlation functions from N-body 
simulation data}

\subsection{Particle distribution on the light cone from N-body simulations
\label{subsec:lcfromnbody}}

%%%% TABLE 1 %%%%%%%%%%%%%%%%%%%%%%%%%%%%%%%%%%%%%%%%%%%%%%%%%%%%%%%%%%%%%%
\begin{table*}
 \caption{Parameters in $N$-body simulations.}
\label{tab:parameters}
\begin{center}
\begin{tabular}{lcccccccc}
\hline
Model & $\Omega_0$ & $\lambda_0$ & $h$ & $\sigma_8$ & Box size &
Force resolution & $z_{\rm max}$ \\
& & & & & $L_X\times L_Y\times L_Z [h^{-3}$Mpc$^3$] &
[$h^{-1}$Mpc]& \\
\hline
SCDM small box & 1 & 0 & 0.5 & 0.6 & $80\times80\times160$ & 0.31 & 0.4 \\
SCDM large box & 1 & 0 & 0.5 & 0.6 & $240\times240\times480$ & 0.94 & 2 \\
$\Lambda$CDM small box & 0.3 & 0.7 & 0.7 & 0.9 &
$120\times120\times240$ & 0.45 & 0.4 \\
$\Lambda$CDM large box & 0.3 & 0.7 & 0.7 & 0.9 &
$360\times360\times620$ & 1.4 & 2 \\
\hline
\end{tabular}
\end{center}
\end{table*} 
%%%%%%%%%%%%%%%%%%%%%%%%%%%%%%%%%%%%%%%%%%%%%%%%%%%%%%%%%%%%%%%%%%%%%%%%%%%%%

%%%% TABLE 2 %%%%%%%%%%%%%%%%%%%%%%%%%%%%%%%%%%%%%%%%%%%%%%%%%%%%%%%%%%%%%%
\begin{table*}
 \caption{Parameter values for the polynomial evolution model of Boyle
 et al. (2000).}
\label{tab:boyle}
\begin{center}
\begin{tabular}{cccccccc}
\hline
 $\Omega_0$ & $\lambda_0$ & $\alpha$ & $\beta$ & $M_{\rm B}^\ast-5
 \log h$ & $k_1$ & $k_2$ & $\Phi^\ast [h^3$Mpc$^{-3}$mag$^{-1}]$\\
\hline
1 & 0 & 3.45 & 1.63 & -20.59 & 1.31 & $-0.26$ & $0.80 \times 10^{-5}$ \\
0.3 & 0.7 & 3.41 & 1.58 & -21.14 & 1.36 & $-0.27$ & $2.88 \times 10^{-6}$ \\
\hline
\end{tabular}
\end{center}
\end{table*} 
%%%%%%%%%%%%%%%%%%%%%%%%%%%%%%%%%%%%%%%%%%%%%%%%%%%%%%%%%%%%%%%%%%%%%%%%%%%%%

%%%%% TABLE 3 %%%%%%%%%%%%%%%%%%%%%%%%%%%%%%%%%%%%%%%%%%%%%%%%%%%%%%%%%%%%%%%
\begin{table*}
 \caption{Summary of number of particles in each realization (real
 space/redshift space).}
\label{tab:particlenumber}
\begin{center}
\begin{tabular}{lccccc}
\hline
Model & Realization & Total & Random selection & LF based selection &
LF based with random selection \\
\hline
%%% small box, B_lim=19
SCDM small box & 1 & 8193106 / 8216016 & 10258 / 10282 & 125477 / 125289 & 8546 / 8540 \\
& 2 & 8291309 / 8311402 & 10388 / 10413 & 168363 / 168999 & 11444 / 11497 \\
& 3 & 8448034 / 8479865 & 10591 / 10627 & 165217 / 165192 & 11221 / 11218 \\
& 4 & 9181442 / 9250736 & 11533 / 11618 & 175769 / 175773 & 11927 / 11927 \\
& 5 & 8263119 / 8324278 & 10340 / 10434 & 178135 / 177348 & 12075 / 12025 \\
%%% large box, B_lim=21
SCDM large box & 1 & 6253827 / 6254790 & 10481 / 10482 & 2037314 / 2041146 & 10348 / 10363 \\
& 2 & 6321816 / 6319899 & 10591 / 10582 & 2077216 / 2077310 & 10552 / 10552 \\
& 3 & 6346239 / 6342617 & 10626 / 10623 & 2090246 / 2090222 & 10622 / 10622 \\
& 4 & 6423700 / 6417089 & 10767 / 10757 & 2102122 / 2099505 & 10671 / 10664 \\
& 5 & 6298022 / 6300195 & 10546 / 10552 & 2077854 / 2079776 & 10553 / 10564 \\
%%% small box, B_lim=19
$\Lambda$CDM small box & 1 & 3253963 / 3224663 & 7589 / 7512 & 43377 /
42960 & 8760 / 8666 \\
& 2 & 4326797 / 4341618 & 10025 / 10050 & 48690 / 48581 & 9808 / 9791 \\
& 3 & 4429032 / 4423464 & 10263 / 10258 & 62073 / 62274 & 12517 / 12553 \\
& 4 & 4859939 / 4842481 & 11245 / 11201 & 59105 / 59022 & 11903 / 11890 \\
& 5 & 4993640 / 4988234 & 11532 / 11517 & 72490 / 72674 & 14608 / 14635 \\
%%% large box, B_lim=21
$\Lambda$CDM large box & 1 & 5358865 / 5370894 & 9834 / 9863 & 1427660
/ 1429062  & 9588 / 9593 \\
& 2 & 5277031 / 5286441 & 9665 / 9681 & 1415712 / 1418226 & 9498 / 9516 \\
& 3 & 5625180 / 5631157 & 10322 / 10326 & 1507183 / 1507424 & 10174 / 10175 \\
& 4 & 5630820 / 5631761 & 10326 / 10326 & 1511963 / 1511565 & 10219 / 10218 \\
& 5 & 5606636 / 5612974 & 10287 / 10300 & 1507176 / 1508490 & 10174 / 10187 \\
\hline
\end{tabular}
\end{center}
\end{table*} 
%%%%%%%%%%%%%%%%%%%%%%%%%%%%%%%%%%%%%%%%%%%%%%%%%%%%%%%%%%%%%%%%%%%%%%%%%%%%%

We test the theoretical modeling against simulation results, we focus
on two spatially-flat cold dark matter models, SCDM and $\Lambda$CDM,
adopting a scale-invariant primordial power spectral index of $n=1$.
Their cosmological parameters are listed in Table
\ref{tab:parameters}.  
While SCDM are known to have several problems
in reproducing the recent observations (e.g., de Bernardis et al.
2000), this model is suitable for testing the theoretical formula
since the clustering evolution on the light-cone is more significant.
We use a series of $N$-body simulations originally constructed for the
study of weak lensing statistics (Hamana et al. 2000, in preparation).
These simulations were generated with a vectorized PM code (Moutarde
at al.~1991) modified to run in parallel on several processors of a
CRAY-98 (Hivon 1995).  They use $256^2\times512$ particles and the
same number of force mesh in a periodic rectangular comoving box.  We
use both the small and large boxes (Table \ref{tab:parameters}).

The initial conditions are generated adopting the transfer function of
Bond \& Efstathiou (1984, see also Jenkins et al.~1998) with the shape
parameter $\Gamma=\Omega_0 h$.  The amplitude of the power
spectrum is normalized by the cluster abundance (Eke, Cole \& Frenk
1996; Kitayama \& Suto 1997).

%%%%%% FIG 2 %%%%%%%%%%%%%%%%%%%%%%%%%%%%%%%%%%%%%%%%%%%%%%%%%%%%
\begin{figure}[ht]
\begin{center}
   \leavevmode \epsfxsize=8.4cm \epsfbox{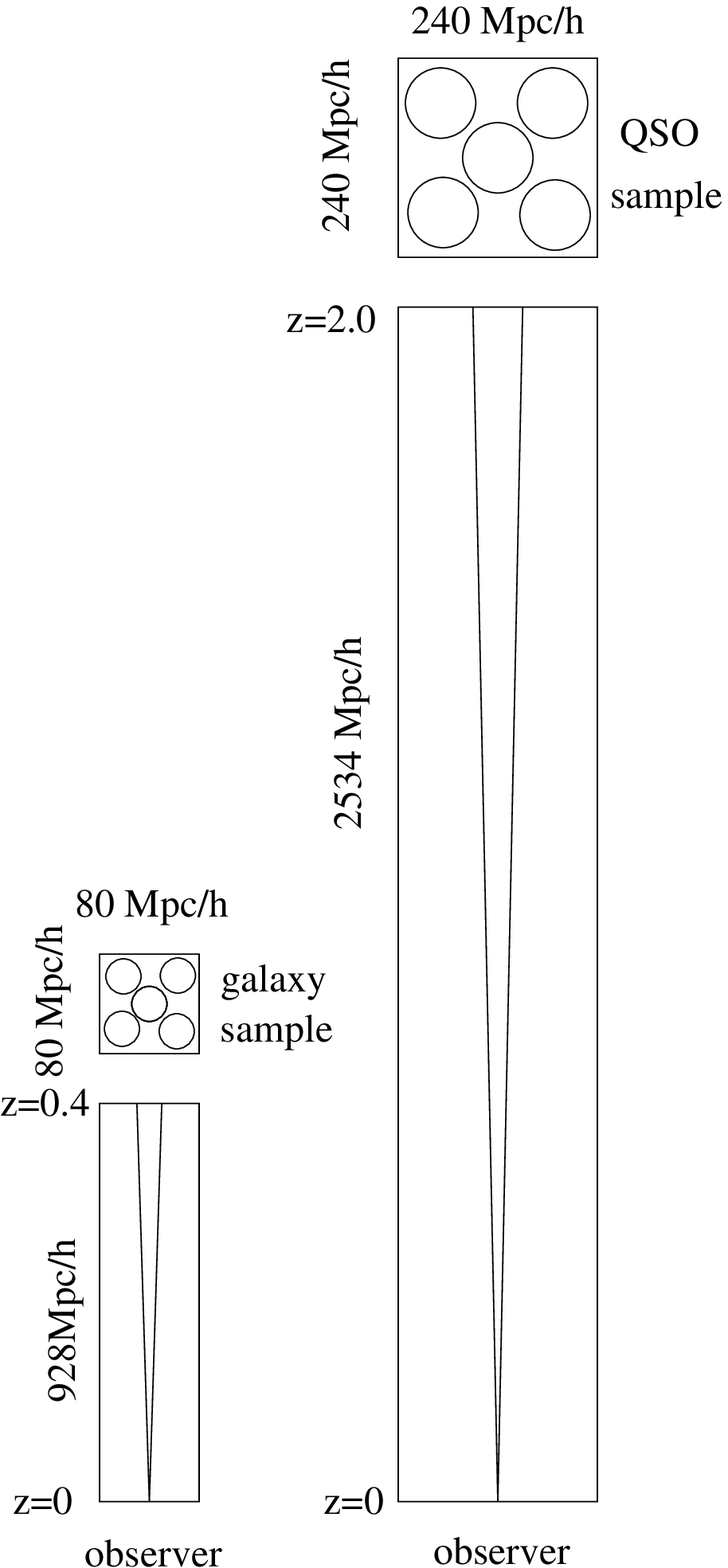}
\end{center}
\caption{Schematic geometry of our light-cone samples. 
The comoving distances denoted in the figure are for SCDM model.
In the case of $\Lambda$CDM model, the redshifts are not changed but
the radial comoving distance is 1086(3626)$h^{-1}$Mpc and the side
length is 120(360)$h^{-1}$Mpc for the small(large) box.}
\label{fig:lightcone}
\end{figure}
%%%%%%%%%%%%%%%%%%%%%%%%%%%%%%%%%%%%%%%%%%%%%%%%%%%%%%%%%%%%%%%%%%%%%

Using the above simulation data, we generated light-cone samples as
follows; first, we adopt a distance observer approximation and assume
that the line-of-sight direction is parallel to $Z$-axis regardless
with its $(X,Y)$ position (Fig.\ref{fig:lightcone}). Second, we
periodically duplicate the simulation box along the $Z$-direction so
that at a redshift $z$, the position and velocity of those particles
locating within an interval $\chi(z) \pm \Delta\chi(z)$ are dumped,
where $\Delta\chi(z)$ is determined by the output time-interval of the
original $N$-body simulation.  Finally we extract five independent
(non-overlapping) cone-shape samples with the angular radius of 1
degree (the field-of-view of $\pi$ degree$^2$), each for small and
large boxes as illustrated in Figure \ref{fig:lightcone}.  In this
manner, we have generated mock data samples on the light-cone
continuously extending up to $z=0.4$ (relevant for galaxy samples) and
$z=2.0$ (relevant for QSO samples), respectively from the small and
large boxes.  While the above procedure selects the same particle at
several different redshifts, this does not affect our conclusion below
because we are mainly interested in scales much below the box size
along the $Z$-direction, $L_Z$.

In practice, we apply the above procedure separately in real and
redshift spaces by using $z_{\rm real}$ and $z_{\rm obs}$ of each
particle (see eq.[\ref{eq:zobs}] below).  The total numbers of
particles in those realizations are listed in Table
\ref{tab:particlenumber}.

\subsection{Pair counts in real and redshift spaces}

Two-point correlation function is estimated by the conventional 
pair-count adopting the estimator proposed by Landy \& Szalay (1993):
%%%%%%%%%%%%%%%%%%%%%%%%%%%%%%%%%%%%%%%%%%%%%%%
\begin{equation}
\xi(x)= \frac{DD(x)-2DR(x)+RR(x)}{RR(x)} .
\end{equation}
%%%%%%%%%%%%%%%%%%%%%%%%%%%%%%%%%%%%%%%%%%%%%%%
For this purpose, we distribute the same number of particles over the
light-cone in a completely random fashion.  When the number of
particles in a realization exceeds $10^6$, we randomly select 10,000
particles as center particles in counting the pairs.  Otherwise we use
all the particles in the pair counts.

The comoving separation $x_{12}$ of two objects located at $z_1$ and
$z_2$ with an angular separation $\theta_{12}$ is given by
%%%%%%%%%%%%%%%%%%%%%%%%%%%%%%%%%%%%%%%%%%%%%%%%%%%%%%%%%%%%%%%%%%%
\begin{eqnarray} 
x_{12}^2 &=& x_1^2 + x_2^2 
-K x_1^2 x_2^2 (1+\cos^2\theta_{12})\nonumber\\
&&-2 x_1 x_2 \sqrt{1-Kx_1^2}\sqrt{1-Kx_2^2}\cos\theta_{12} ,
\end{eqnarray}
%%%%%%%%%%%%%%%%%%%%%%%%%%%%%%%%%%%%%%%%%%%%%%%%%%%%%%%%%%%%%%%%%%%
where $x_1\equiv D_c(z_1)$ and $x_2\equiv D_c(z_2)$. 

In redshift space, the observed redshift $z_{\rm obs}$ for each object
differs from the ``real'' one $z_{\rm real}$ due to 
the velocity distortion effect:
%%%%%%%%%%%%%%%%%%%%%%%%%%%%%%%%%%%%%%%%%%%%%%%%%%%%%%%%%%%%%%%%%%%
\begin{eqnarray} 
\label{eq:zobs}
z_{\rm obs} = z_{\rm real} + (1+z_{\rm real}) v_{\rm pec} ,
\end{eqnarray}
%%%%%%%%%%%%%%%%%%%%%%%%%%%%%%%%%%%%%%%%%%%%%%%%%%%%%%%%%%%%%%%%%%%
where $v_{\rm pec}$ is the line of sight relative peculiar velocity
between the object and the observer in {\it physical} units. Then the
comoving separation $s_{12}$ of two objects in redshift space is
computed as
%%%%%%%%%%%%%%%%%%%%%%%%%%%%%%%%%%%%%%%%%%%%%%%%%%%%%%%%%%%%%%%%%%%
\begin{eqnarray} 
s_{12}^2 &=&s_1^2 + s_2^2 
-K s_1^2 s_2^2 (1+\cos^2\theta_{12})\nonumber\\
&&-2 s_1 s_2 \sqrt{1-Ks_1^2}\sqrt{1-Ks_2^2}\cos\theta_{12} ,
\end{eqnarray}
%%%%%%%%%%%%%%%%%%%%%%%%%%%%%%%%%%%%%%%%%%%%%%%%%%%%%%%%%%%%%%%%%%%
where $s_1\equiv D_c(z_{\rm obs, 1})$ and $s_2\equiv D_c(z_{\rm obs,
  2})$.

\subsection{Selection functions}

In properly predicting the power spectra on the light cone, the
selection function should be specified. In this subsection, we describe
the selection functions appropriate for galaxies and quasars samples.

\begin{figure}[t]
\begin{center}
   \leavevmode \epsfysize=9cm \epsfbox{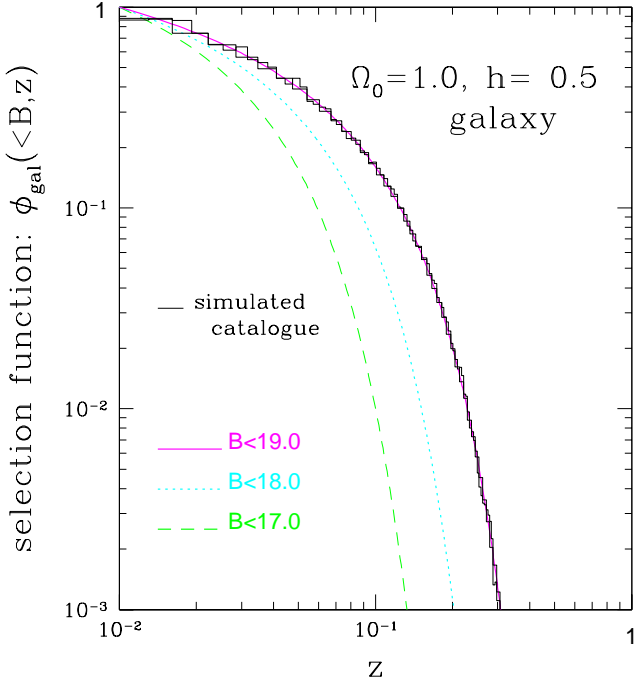}
\end{center}
\caption{ Selection function of galaxies in a case of SCDM model on
  the basis of the B-band luminosity function of APM galaxies (Loveday
  et al. 1992).}
\label{fig:galselfunc}
\end{figure}

For galaxies, we adopt a B-band luminosity function of the APM galaxies
(Loveday et~al. 1992) fitted to the Schechter function:
%%%%%%%%%%%%%%%%%%%%%%%%%%%%%%%%%%%%%%%%%%%%%%%%%%%%%%%%%%%%%%
\begin{equation}
  \phi(L)dL=\phi^*\biggl({L\over L^*}\biggr)^\alpha
  \exp\biggl(-{L\over L^*}\biggr)d\biggl({L\over L^*}\biggr), 
\end{equation}
%%%%%%%%%%%%%%%%%%%%%%%%%%%%%%%%%%%%%%%%%%%%%%%%%%%%%%%%%%%%%%
with $\phi^*=1.40\times 10^{-2} h^3{\rm Mpc}^{-3}$, $\alpha=-0.97$, and
$M_{B}^*=-19.50+5\log_{10} h$.  Then the comoving number density of
galaxies at $z$ which are brighter than the limiting magnitude $B_{\rm
lim}$ is given by
%%%%%%%%%%%%%%%%%%%%%%%%%%%%%%%%%%%%%%%%%%%%%%%%%%%%%%%%%%%%%%%
\begin{eqnarray}
  n_{\rm gal}(z,<B_{\rm lim})&=&\int_{L(B_{\rm lim},z)}^\infty \phi(L)
  dL\nonumber\\
  &=&\phi^* \Gamma[(\alpha+1,x(B_{\rm lim},z)] ,
\end{eqnarray}
%%%%%%%%%%%%%%%%%%%%%%%%%%%%%%%%%%%%%%%%%%%%%%%%%%%%%%%%%%%%%%%
where 
%%%%%%%%%%%%%%%%%%%%%%%%%%%%%%%%%%%%%%%%%%%%%%%%%%%%%%%%%%%%%%%%%%%%%%%%%%
\begin{equation}
  x(B_{\rm lim},z) \equiv {L(B_{\rm lim},z) \over L^*}
= \left[{d_L(z)\over 1 h^{-1}{\rm Mpc}}\right]^2 10^{2.2-0.4 B_{\rm lim}},
\end{equation}
%%%%%%%%%%%%%%%%%%%%%%%%%%%%%%%%%%%%%%%%%%%%%%%%%%%%%%%%%%%%%%%%%%%%%%%%%%
and $\Gamma[\nu,x]$ is the incomplete Gamma function.
Figure \ref{fig:galselfunc} plots the selection function defined by
%%%%%%%%%%%%%%%%%%%%%%%%%%%%%%%%%%%%%%%%%%%%%%%%%%%%%%%%%%%%%%%
\begin{eqnarray}
\phi_{\rm gal}(<B_{\rm lim},z)
\equiv {n_{\rm gal}(z,<B_{\rm lim}) \over 
n_{\rm gal}(z_{\rm min},<B_{\rm  lim})} 
\end{eqnarray}
%%%%%%%%%%%%%%%%%%%%%%%%%%%%%%%%%%%%%%%%%%%%%%%%%%%%%%%%%%%%%%%
with $z_{\rm min}=0.01$. 

For quasars, we adopt the B-band luminosity recently determined by
Boyle et al. (2000) from the 2dF QSO survey data:
%%%%%%%%%%%%%%%%%%%%%%%%%%%%%%%%%%%%%%%%%%%%%%%%%%%%%%%%%%%%%%%%%%%
\begin{eqnarray} 
\Phi(M_{\rm B},z) = {\Phi^* \over
10^{0.4(1-\alpha)[M_{\rm B}-M_{\rm B}^*(z)]} + 10^{0.4(1-\beta)[M_{\rm
B}-M_{\rm B}^*(z)]} } . 
\label{eq:phiqso} 
\end{eqnarray}
%%%%%%%%%%%%%%%%%%%%%%%%%%%%%%%%%%%%%%%%%%%%%%%%%%%%%%%%%%%%%%%%%%%
In the case of the polynomial evolution model:
%%%%%%%%%%%%%%%%%%%%%%%%%%%%%%%%%%%%%%%%%%%%%%%%%%%%%%%%%%%%%%%%%%%%%%%%%%
\begin{equation}
M_{\rm B}^*(z) = M_{\rm B}^*(0) - 2.5 (k_1 z + k_2 z^2) ,
\end{equation}
%%%%%%%%%%%%%%%%%%%%%%%%%%%%%%%%%%%%%%%%%%%%%%%%%%%%%%%%%%%%%%%%%%%%%%%%%%
and we adopt the sets of their best-fit parameters listed in Table
\ref{tab:boyle} for our SCDM and $\Lambda$CDM.

To compute the B-band apparent magnitude from a quasar of absolute
magnitude $M_{\rm B}$ at $z$ (with the luminosity distance
$d_{\rm L}(z)$), we applied the K-correction:
%%%%%%%%%%%%%%%%%%%%%%%%%%%%%%%%%%%%%%%%%%%%%%%%%%%%%%%%%%%%%%%%%%%
\begin{equation}
  B = M_{\rm B} + 5 \log(d_{\rm L}(z)/ 10 {\rm pc}) 
- 2.5(1-p)\log (1+z) 
\end{equation}
%%%%%%%%%%%%%%%%%%%%%%%%%%%%%%%%%%%%%%%%%%%%%%%%%%%%%%%%%%%%%%%%%%%
for the quasar energy spectrum $L_\nu \propto \nu^{-p}$ (we use
$p=0.5$). 

\begin{figure}[t]
\begin{center}
   \leavevmode \epsfysize=9cm \epsfbox{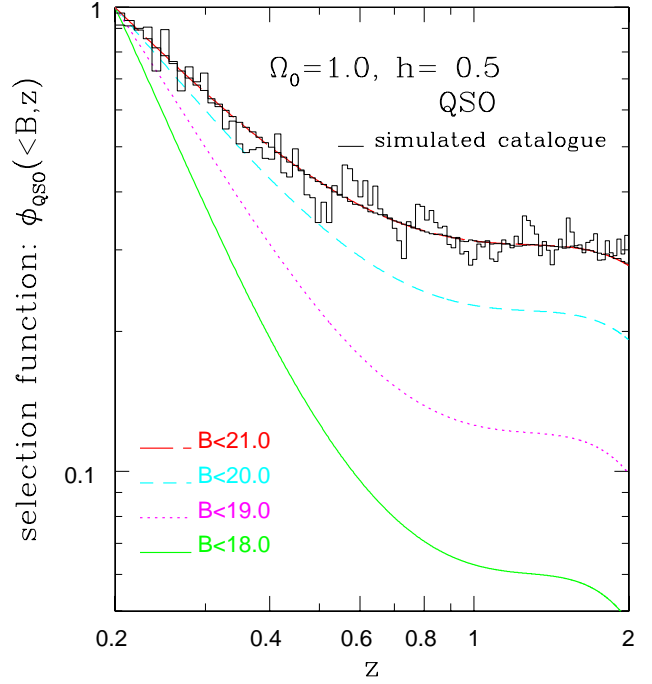}
\end{center}
\caption{ Selection function of QSOs in a case of SCDM model on the
basis of the 2dF QSO sample (Boyle et al. 2000).}
\label{fig:qsoselfunc}
\end{figure}

Then the comoving number density of
QSOs at $z$ which are brighter than the limiting magnitude $B_{\rm
lim}$ is given by
%%%%%%%%%%%%%%%%%%%%%%%%%%%%%%%%%%%%%%%%%%%%%%%%%%%%%%%%%%%%%%%
\begin{eqnarray}
  n_{\rm QSO}(z,<B_{\rm lim})=\int_{-\infty}^{M(B_{\rm lim},z)}
\Phi(M_{\rm B},z) dM_{\rm B}.
\end{eqnarray}
%%%%%%%%%%%%%%%%%%%%%%%%%%%%%%%%%%%%%%%%%%%%%%%%%%%%%%%%%%%%%%%
Figure \ref{fig:qsoselfunc} plots the selection function defined by
%%%%%%%%%%%%%%%%%%%%%%%%%%%%%%%%%%%%%%%%%%%%%%%%%%%%%%%%%%%%%%%
\begin{eqnarray}
\phi_{\rm QSO}(<B_{\rm lim},z)
\equiv {n_{\rm QSO}(z,<B_{\rm lim}) \over 
n_{\rm QSO}(z_{\rm min},<B_{\rm  lim})} 
\end{eqnarray}
%%%%%%%%%%%%%%%%%%%%%%%%%%%%%%%%%%%%%%%%%%%%%%%%%%%%%%%%%%%%%%%
with $z_{\rm min}=0.2$. 

%%%%%% FIG 3 %%%%%%%%%%%%%%%%%%%%%%%%%%%%%%%%%%%%%%%%%%%%%%%%%%%%%%%%%
%%%%%%%%%%%%%%%%%%%%%%%%%%%%%%%%%%%%%%%%%%%%%%%%%%%%%%%%%%%%%%%%%%%%%

%%%%%%% FIG 4 %%%%%%%%%%%%%%%%%%%%%%%%%%%%%%%%%%%%%%%%%%%%%%%%%%%%%%%%%
%%%%%%%%%%%%%%%%%%%%%%%%%%%%%%%%%%%%%%%%%%%%%%%%%%%%%%%%%%%%%%%%%%%%%

In practice, we adopt the galaxy selection function $\phi_{\rm
gal}(<B_{\rm lim},z)$ with $B_{\rm lim}=19$ and $z_{\rm min}=0.01$ for
the small box realizations, while the QSO selection function
$\phi_{\rm QSO}(<B_{\rm lim},z)$ with $B_{\rm lim}=21$ and $z_{\rm
min}=0.2$ for the large box realizations.  We do not introduce the
spatial biasing between selected particles and the underlying dark
matter, which will be discussed elsewhere.  For comparison, we also
select the similar number of particles randomly but independently of
their redshifts.  It should be emphasized here that our simulated data
are constructed to match the {\it shape} of the above selection
functions but {\it not} the amplitudes of the number densities. The
field-of-view of our simulated data, $\pi$ degree$^2$, is
substantially smaller than those of 2dF and SDSS, and we sample
particles much more densely than the realistic number density. Since
our main purpose of this paper is to test the reliability of the
theoretical modeling described in section 2, and {\it not} to present
detailed predictions, this does not change our conclusions below.  The
numbers of the selected particles in each realization are listed in
Table \ref{tab:particlenumber}. The averaged {\it selection functions}
for our five realizations in real space are plotted as histograms in
Figures \ref{fig:galselfunc} and \ref{fig:qsoselfunc}.

\section{Results}

%%%%%%% FIG 5 & 6 %%%%%%%%%%%%%%%%%%%%%%%%%%%%%%%%%%%%%%%%%%%%%%%%%%%%%%%
\begin{figure*}
\begin{center}
   \leavevmode \epsfxsize=15cm \epsfbox{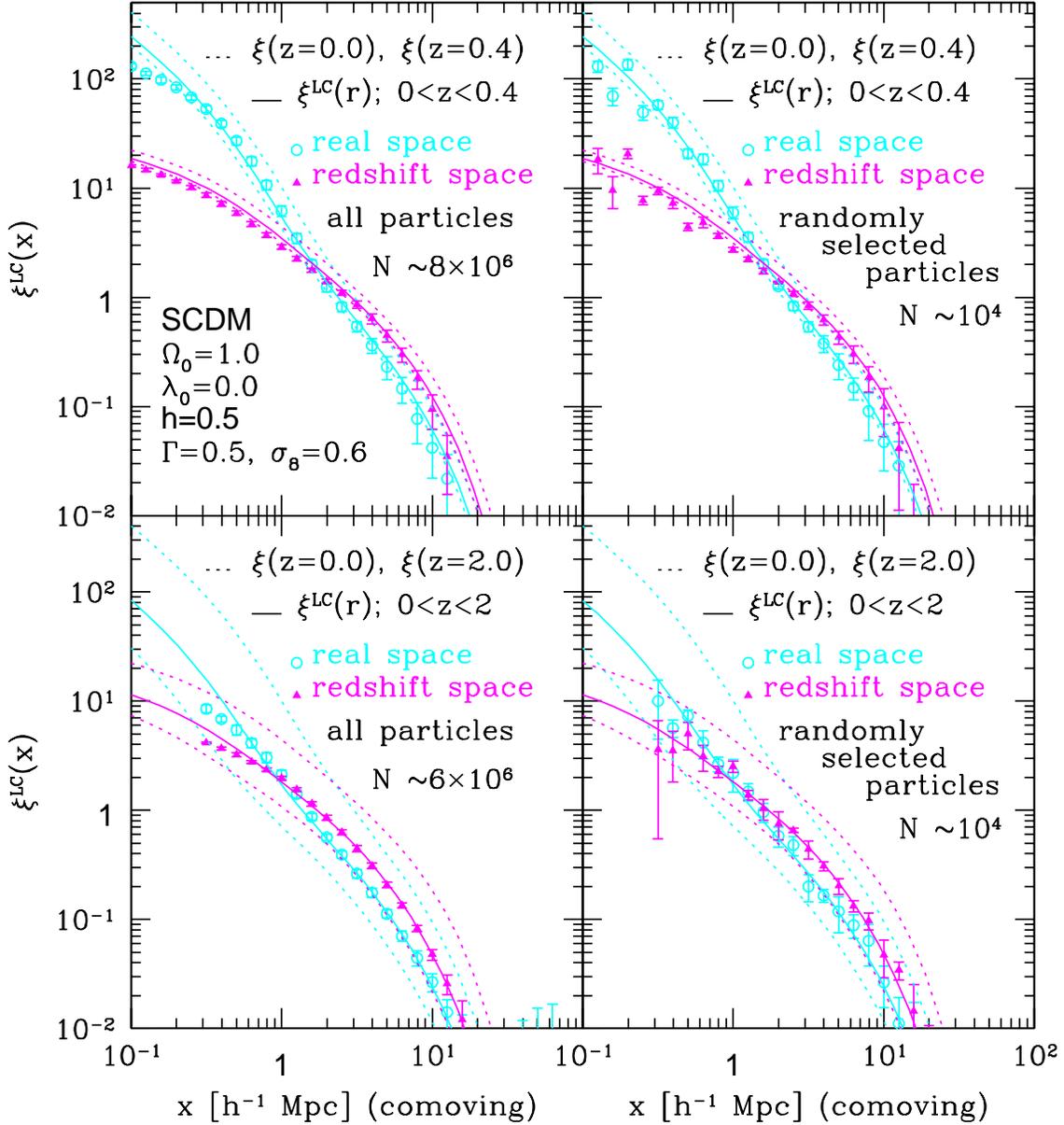}
\end{center}
\caption{ Mass two-point correlation functions on the light cone
without redshift-dependent selection functions in SCDM model.
{\it Upper:} $z<0.4$, {\it Lower:} $0<z<2.0$.  {\it Left:} all particles
on the light cone, {\it Right:} randomly
selected particles.}
\label{fig:xilc_mass}
\end{figure*}
%%%%%
\begin{figure*}
\begin{center}
   \leavevmode \epsfxsize=15cm \epsfbox{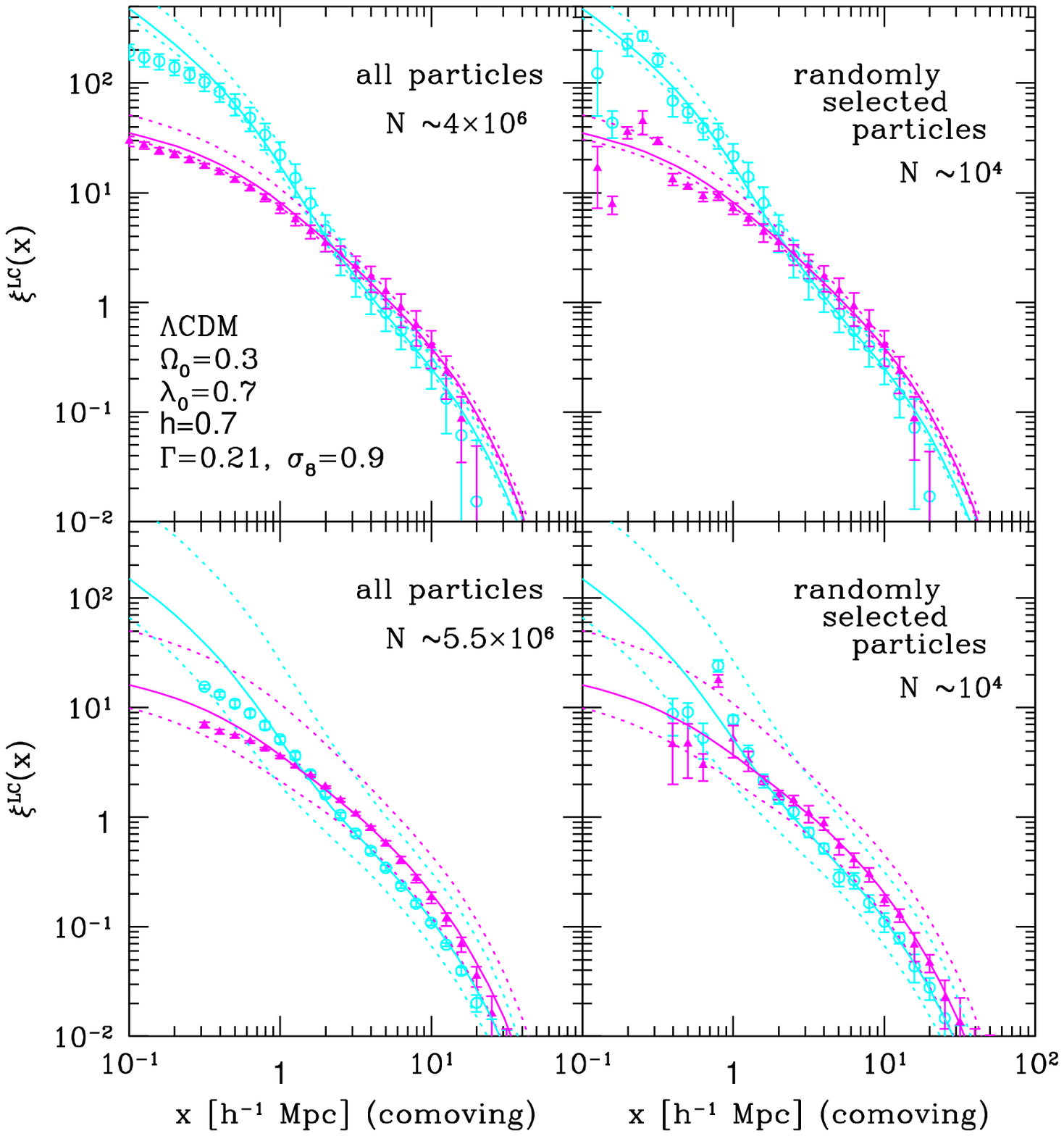}
\end{center}
\caption{  Same as Figure \ref{fig:xilc_mass} but for $\Lambda$CDM
model.}
\label{fig:xilc_mass_lcdm}
\end{figure*}
%%%%%%%%%%%%%%%%%%%%%%%%%%%%%%%%%%%%%%%%%%%%%%%%%%%%%%%%%%%%%%%%%%%%%

Consider first the two-point correlation functions for particles on
the light cone but without redshift-dependent selection.  Figures
\ref{fig:xilc_mass} and \ref{fig:xilc_mass_lcdm} plot those
correlations for $z<0.4$ samples from small-box simulations (upper
panels) and for $z<2$ from large-box ones (lower panels), for SCDM and
$\Lambda$CDM, respectively.  In these figure, we plot the averages over
the five realizations (Table \ref{tab:particlenumber}) in open circles
(real space) and in solid triangles (redshift space), and the quoted
error-bars represent the standard deviation among them.  If we use all
particles from simulations (left panels), the agreement between the
theoretical predictions (solid lines) and simulations (symbols) is
quite good.  The scales where the simulation data in real space become
smaller than the corresponding theoretical predictions simply reflect
the force resolution of the simulations listed in Table \ref{tab:parameters}.

In order to examine the robustness of the estimates from the simulated
data, we randomly selected $N \sim 10^4$ particles from the entire
light-cone volume (independently of their redshifts). The resulting
correlation functions are plotted in the right panels. It is
remarkable that the estimates on scales larger than $\sim 1 h^{-1}{\rm
Mpc}$ are almost the same. This also indicates that the error-bars in
our data are dominated by the sample-to-sample variation among the
different realizations.

%%%%% FIG 7 & 8 %%%%%%%%%%%%%%%%%%%%%%%%%%%%%%%%%%%%%%%%%%%%%%%%%%%%%%%
\begin{figure*}
\begin{center}
   \leavevmode \epsfxsize=15cm \epsfbox{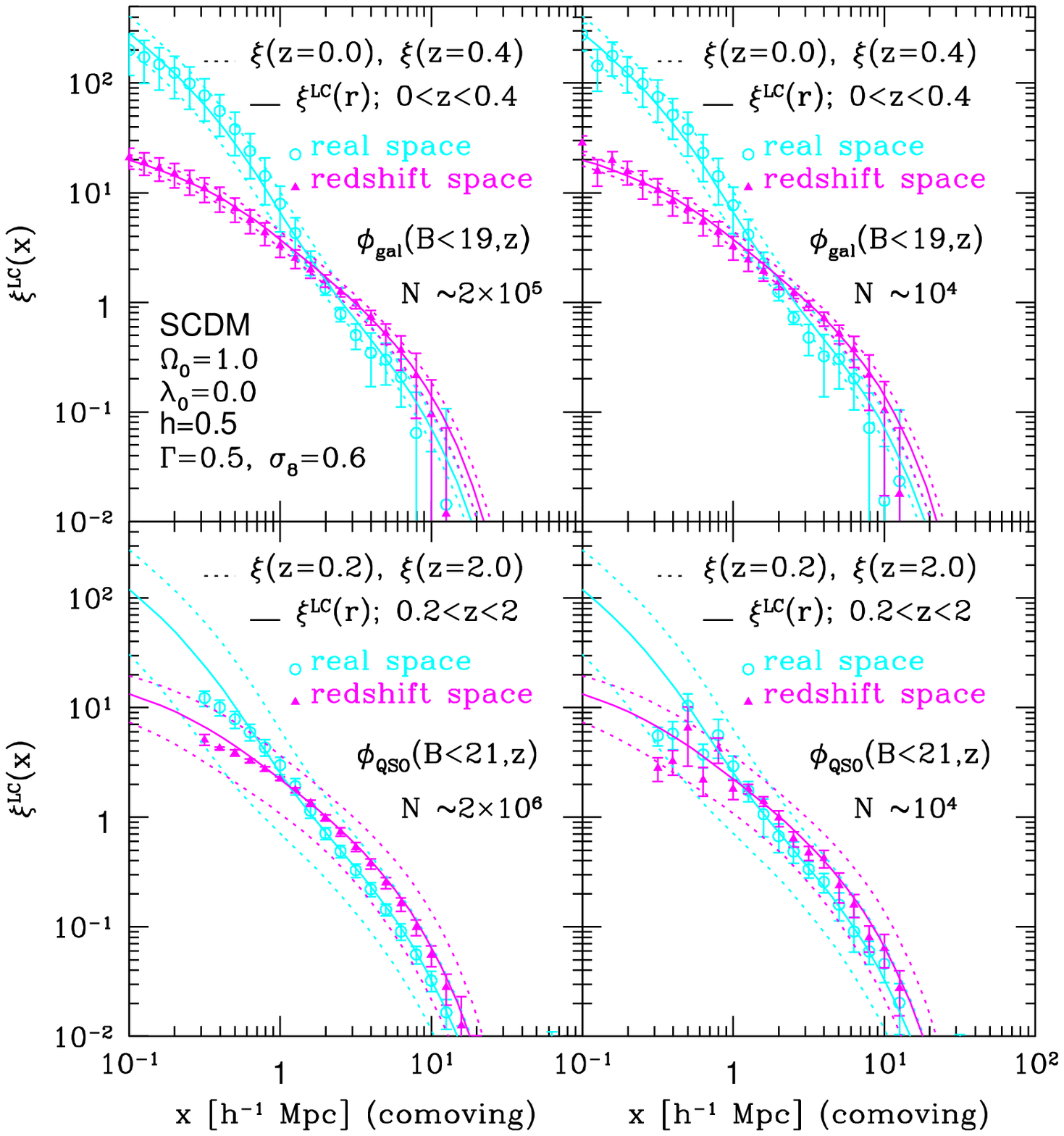}
\end{center}
\caption{ Mass two-point correlation functions on the light cone for
 particles with redshift-dependent selection functions in SCDM model.
{\it Upper:} $z<0.4$, {\it Lower:} $0.2<z<2.0$.
{\it Left:} with selection function whose shape
 is the same as that of the B-band magnitude limit of 
19 for galaxies (upper) and 21 for QSOs (lower). 
{\it Right:} randomly selected $N\sim 10^4$ particles from the
 particles in the left results.}
\label{fig:xilc_phi}
\end{figure*}
%%%%%%
\begin{figure*}
\begin{center}
   \leavevmode \epsfxsize=15cm \epsfbox{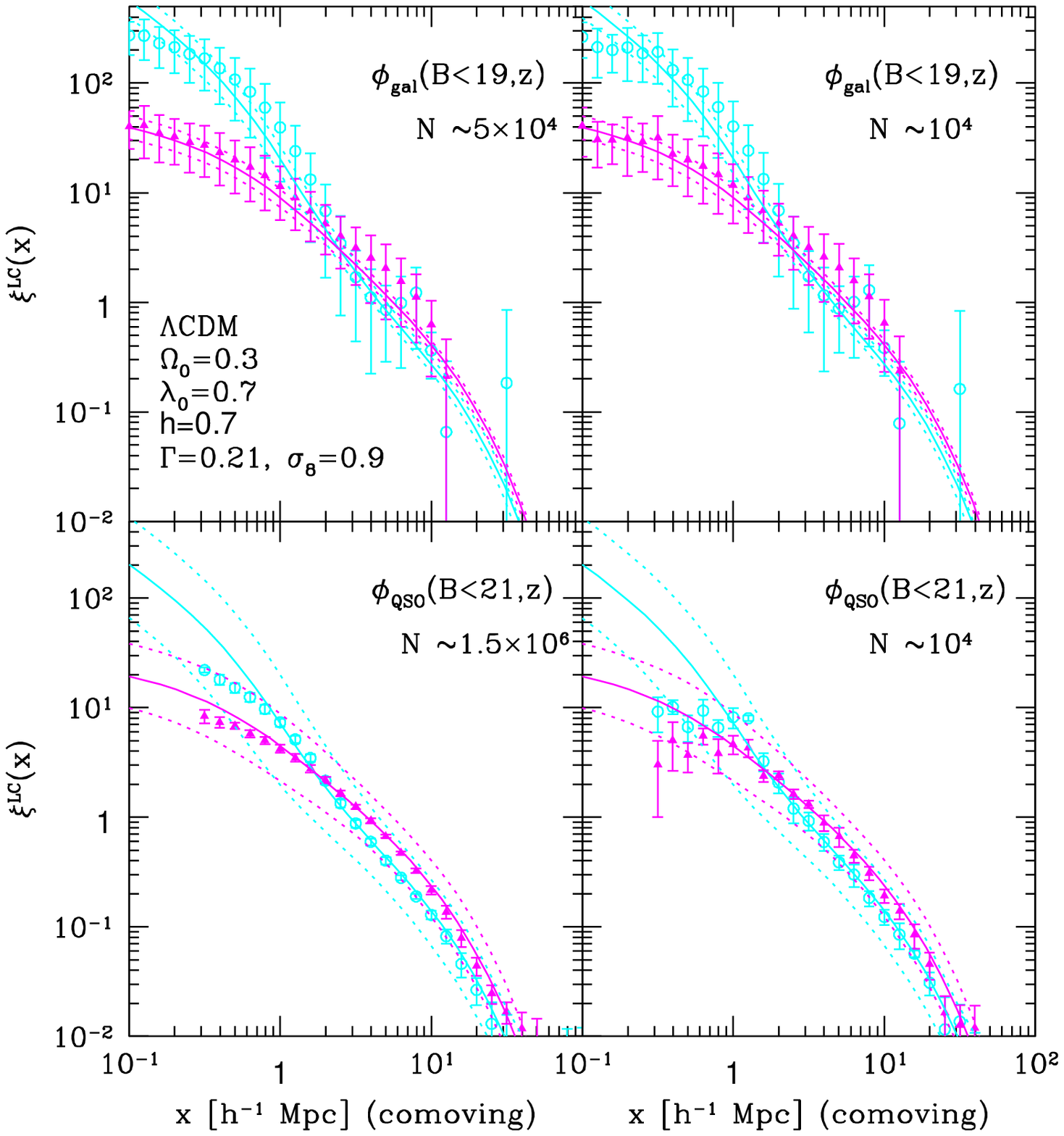}
\end{center}
\caption{ Same as Figure \ref{fig:xilc_phi} but for $\Lambda$CDM
model.}
\label{fig:xilc_phi_lcdm}
\end{figure*}
%%%%%%%%%%%%%%%%%%%%%%%%%%%%%%%%%%%%%%%%%%%%%%%%%%%%%%%%%%%%%%%%%%%%%

Next we examine the effect of selection functions.  Figures
\ref{fig:xilc_phi} and \ref{fig:xilc_phi_lcdm} plot the two-point
correlation functions in SCDM and $\Lambda$CDM, respectively, taking
account of the selection functions described in the subsection 3.3.
It is clear that the simulation results and the predictions are in
good agreement. It should be noted that the results shown in the
upper-left panel (intended to correspond to {\it galaxies}) have
substantially larger error-bars compared with the corresponding ones
in Figures \ref{fig:xilc_mass} and \ref{fig:xilc_mass_lcdm}. This is
an artifact to some extent because of the very small survey volume in
our light-cone samples; if one applies the galaxy selection function
which rapidly decreases as $z$ (see, Fig.\ref{fig:galselfunc}), the
resulting structure mainly probes the universe at $z < 0.1$ and thus
large-scale nonlinearity or variation for the different line-of-sight
becomes significant. If we are able to use the same number of
particles but extending over the much larger volume, the
sample-to-sample variations should be substantially smaller. This
interpretation is supported by the upper-right panel where we randomly
sample $N\sim 10^4$ particles from those used in the upper-left panel.
Despite the fact that the number of particles is only 5\% (20\%) of
the original one for SCDM ($\Lambda$CDM) model, the resulting
correlation functions and their error-bars remain almost unchanged.
The lower panels corresponding to {\it QSOs} show the similar trend.

\section{Conclusions and discussion}

We have presented detailed comparison between the theoretical modeling
and the direct numerical results of the two-point correlation
functions on the light-cone. In short, we have quantitatively shown
that the previous theoretical models by Yamamoto \& Suto (1999) and
Yamamoto, Nishioka \& Suto (1999) are quite accurate on scales $1
h^{-1} {\rm Mpc} < x < 20 h^{-1} {\rm Mpc}$ where the numerical
simulations are reliable. It is also encouraging that this conclusion
remains true even for the particle number of around $10^4$.  In fact,
the error-bars in our estimates of the two-point correlation functions
are dominated by the sample-to-sample variance due to the limited
angular-size ($\pi$-degree$^2$) and thus the limited volume.

In order for the more realistic evaluation of the statistical and
systematic uncertainties, one needs mock light-cone data samples with
a much wider sky coverage.  More importantly such datasets enable one
to access the effect of biasing on the two-point correlation functions
on the light-cone. Since our present study indicated that all the
physical effects except for the biasing are well described by the
existing theoretical models, it is very interesting to examine in
detail how to extract the effect of the galaxy/QSO biasing from the
upcoming redshift survey on the basis of the above mock samples.  We
plan to come back to these issues with larger simulation datasets in
near future.

\begin{acknowledgements}
This research was supported in part by the Direction de la Recherche
du Minist{\`e}re Fran{\c c}ais de la Recherche and the Grant-in-Aid by
the Ministry of Education, Science, Sports and Culture of Japan
(07CE2002) to RESCEU.  The computational resources (CRAY-98) for the
present numerical simulations were made available to us by the
scientific council of the Institut du D\'eveloppement et des
Ressources en Informatique Scientifique (IDRIS).

\end{acknowledgements}

%%%%%%%%%%%%%%%%%%%%%%%%%%%%%%%%%%%%%%%%%%%%%%%%%%%%%%%%%%%%%%%%%%%%%%%

%%%%%%%%%%%%%%%%%%%%%%%%%%%%%%%%%%%%%%%%%%%%%%%%%%%%%%%%%%%%%%%%%%%%%%%


\begin{thebibliography}{}
%\parskip=-1pt
%\baselineskip=14pt

\bibitem[Bond \& Efstathiou(1984)]{BE} Bond, J.R., Efstathiou, G., 1984, 
ApJ 285, L45
\bibitem[Boyle et al.(2000)]{Boyle} Boyle, B.J., Shanks, T., Croom,
  S.M., Smith, R.J., Miller, L., Loaring, N., \& Heymans, C. 2000,
  astro-ph/0005368
\bibitem[de Bernardis et al.(2000)]{Bernardis}
de Bernardis, P., et al., 2000, Nature 404, 955
\bibitem[Eke et al.(1996)]{ECF96}Eke, V.R. Cole, S., \& Frenk,
C.S. 1996, MNRAS, 282, 263
\bibitem[Hivon et al.(1995)]{Hivon}
Hivon, E., 1995, PhD thesis, University Paris XI
\bibitem[Jenkins et al.(1998)]{Jenkins}
Jenkins, A.,  et al., 1998, ApJ, 499, 20
\bibitem[Kaiser87]{K87} Kaiser, N. 1987, 227, 1
\bibitem[Kitayama \& Suto(1997)]{KS97}Kitayama, T., \& Suto,
Y. 1997, ApJ, 490, 557
\bibitem[Landy \& Szaly (1993)]{LS93}
Landy, S.D., \& Szalay, A.S., 1993, ApJ, 412, 64
\bibitem[Magira et al.(2000)]{Magira} Magira, H., Jing, Y. P., \&
  Suto, Y. 2000, ApJ, 528, 30
\bibitem[Matarrese et al.(1977)]{Matarresse} Matarrese, S., Coles, P.,
  Lucchin, F., \& Moscardini, L. 1997, MNRAS, 286, 115
\bibitem[Matsubara \& Suto(1996)]{ms96} Matsubara, T., \& Suto, Y.
  1996, ApJ, 470, L1
\bibitem[Matsubara et al.(1997)]{mss97} Matsubara, T., Suto, Y., \&
  Szapudi, I. 1997, ApJ, 491, L1
\bibitem[Mo, Jing, \& B\"orner(1997)]{mjb} Mo, H. J., Jing, Y. P., \&
  B\"orner, G. 1997, MNRAS, 286, 979 (MJB)
\bibitem[Moscardini et al.(1998)]{Moscardini} Moscardini, L., Coles,
  P., Lucchin, \& F., Matarrese, S. 1998, MNRAS, 299, 95
\bibitem[Moutarde et al.(1991)]{Moutarde}
Moutarde, F., Alimi, J.~M., Bouchet, F.~R., Pellat, R., \& Ramani, A.,
1991, ApJ, 382, 377 
\bibitem[Nakamura et al.(1998)]{Nakamura} Nakamura, T. T., Matsubara,
  T., \& Suto, Y. 1998, ApJ, 494, 13
\bibitem[Peacock \& Dodds(1996)]{PD1996} Peacock, J.A., \& Dodds, S.J.
  1996, MNRAS, 280, L19
\bibitem[Suto et al.(1999)]{suto99} Suto, Y., Magira, H., Jing, Y. P.,
  Matsubara, T., \& Yamamoto, K. 1999, Prog.Theor.Phys.Suppl., 133, 183
\bibitem[Suto, Magira, \& Yamamoto(2000)]{suto00} Suto, Y., Magira, H., 
 \& Yamamoto, K. 2000, PASJ, 52, 249
\bibitem[Yamamoto et al.(1999)]{yns99} Yamamoto, K., Nishioka, H., \&
  Suto, Y. 1999, ApJ, 527, 488
\bibitem[Yamamoto \& Suto(1999)]{ys99} Yamamoto, K., \& Suto, Y.
  1999, ApJ, 517, 1
\end{thebibliography}
\end{document}